# Dynamic Correlation Length Growth in Superspin Glass: Bridging Experiments and Simulations

S. Nakamae[1,*], C. Crauste-Thibierge[1,†], D. L'Hôte[1], E.Vincent[1],
E. Dubois[2], V. Dupuis[2] and R. Perzynski[2]

[1]Service de Physique de l'Etat Condensé (CNRS URA 2464) DSM/IRAMIS, CEA Saclay, 91191 Gif sur Yvette, France

[2]Physicochimie des Electrolytes, Colloïdes et Sciences Analytiques, (CNRS UMR 7195) Université Pierre et Marie Curie, 4 Place Jussieu, 75252 Paris, France

Interacting magnetic nanoparticles display a wide variety of magnetic behaviors that are now being gathered in the emerging field of 'supermagnetism.' We have investigated how the out-of-equilibrium dynamics in the disordered superspin glass (SSG) state of a frozen ferrofluid sample is affected by texturation. Via magnetization relaxation experiments at low temperatures, we were able to estimate superspin correlation lengths for both textured and non-textured samples. The comparison with simulations and experiments on atomic spin glasses shows that the dynamic correlations in SSG's appear to develop in a way reminiscent to those in atomic spin glasses at intermediate time/length scales.

## I. INTRODUCTION

Interacting, single-domain ferro(ferri)magnetic nanoparticles (*np*) in solid media (*e.g.,* frozen ferrofluid) are known to undergo a superparamagnetic (SPM)-to-superspin glass (SSG) transition at low temperature.[1,2] The name 'superspin' reflects the large magnetic moment associated with each nanoparticle. Superspins are generally ascribed a strong uni-axial anisotropy energy that results in a dramatic thermally activated increase of their individual flipping time (compared to atomic spins). These 'slow' superspins are thus good candidates for revisiting some of the unsolved questions in the physics of spin glasses (SG) at intermediate time/length scales which were inaccessible by numerical simulations and experiments. Spin glasses, like all other glassy systems, are characterized by the out-of-equilibrium dynamics that fails to establish long-range ordered state due to frozen-in disorders. Instead, magnetic moments slowly establish microscopic local equilibrium whose domain is defined by the correlation lengths. The question on how correlation length ($\xi$) grows in spin-glasses and how it compares to numerical simulations were never clearly answered. The main obstacles were; 1) the experimentally accessible time scale ($10^{-3}$ to $10^5$ sec) in atomic SG's[3-5] is many orders of magnitude larger than what explored by simulations[5-7] and 2) most numerical simulations were made on *Ising* SG's while a vast majority of experiments were performed on *Heisenberg-like* SG's.

---

[*] Corresponding author
[†] Current address : Laboratoire de Physique de l'Ecole Normale Supérieure de Lyon, France

It has been shown numerically by Berthier *et al.*, that the correlation length dynamics of Ising and Heisenberg spin glasses follow different laws[6,7]. A power law behavior, $\xi(t, T) = A(t/\tau_o)^{z(T/T_g)}$ and thus a scaling with $[T/T_g ln(t_w/\tau_o)]$ can reasonably describe the simulation results in *Ising* SG's even at large times, whereas clear deviations from such a scaling occur in *Heisenberg* SG's at $T < T_g$.[7] Here, $A$ is a constant of the order 1, $t$ the lab time, $\tau_o$ the attempt time of a single spin, $z$ dynamic exponent and $t_w$ is the time the system has spent in the SG state. In experiments on real sping lasses, however, such a scaling behavior has only been found in *Heisenberg* SG's[3-5]. To our knowledge there have been two recent studies that attempted to close the time scale gap between numerical simulations and experiments in spin glasses. On one hand, Belletti *et al.* have succeeded in conducting numerical simulation of $\xi$ in *Ising* SG's over a time spanning 11 orders of magnitude[8]. Their simulation yielded correlation lengths that are in the same order of magnitude with the experimentally determined values of *Heisenberg* spin glasses, rather than those of *Ising* SG's. On the other hand, Wood[9] has examined $\xi(t, T)$ from various experimental results on thin-film (2D) and bulk (3D) spin glasses and compared these values to the simulation made by Kisker.[10] When the correlation lengths in thin-film SG's were all fixed to 1.8 times the sample thickness, striking agreement was found between correlation lengths in thin film (susceptibility measurements) and bulk (thermoremanent magnetization measurements) *Heisenberg* spin glass samples and the simulation based on the *Ising* spin glass model.

Maghemite frozen ferrofuids have been found to exhibit Heisenberg SG-like behavior when anisotropy axes are randomly distributed.[11-13] Interestingly, once the anisotropy-axes are uniformly aligned, more Ising SG-like features were observed.[14] Combined with the longer 'flip-time' of individual superspins (~$10^{-9}$ sec at room temperature for ~10 nm *np*'s compared to $10^{-12}$ sec for atomic spins), concentrated ferrofluids (FF) may allow a more direct comparison between real three dimensional Heisenberg-like and Ising-like (super)spin glass systems to their corresponding numerical simulation results.

In this study, the number of correlated superspins $N_s$ was extracted via zero-field-cooled-magnetization (ZFCM) relaxation measurements (see further below for experimental protocol) in the SSG state of two types of frozen ferrofluids; namely, textured and non-textured FF's. In the non-textured FF, both the position and the anisotropy-axis orientation of nanoparticles were kept random, whereas in the textured FF, the particles' anisotropy axes were all aligned by application of a strong magnetic field (1.5 T) in the high temperature liquid state of the carrier fluid. Hereafter, these samples are called "random" and "aligned" samples. The corresponding values of $\xi(t, T)$ were then deduced from $N_s$ using the results from numerical simulations[6,7] on the fractal growth of the correlation in the SSG state.

## II. EXPERIMENTAL AND DATA ANALYSIS METHODS

The glycerol based ferrofluid used in this study was made of maghemite, $\gamma$-Fe$_2$O$_3$, nanoparticles (~8.6 nm diameter) with a ~15% volume fraction. The nanoparticles are magnetically single-domain, possessing an average permanent magnetic moment of ~$10^4$ $\mu_B$ and the anisotropy energy of $E_a$ ~ 640 K[15]. The sample synthesis technique[16] and the texturing procedure[14] (anisotropy axis-alignment at high temperature, high field) are described elsewhere. All magnetization measurements were performed with a commercial SQUID magnetometer (Cryogenic S600). The existence of low temperature SSG state in these ferrofluids was verified via the critical slowing down behavior near the transition temperature, $T_g$ (67~70 K for both) and the ZFCM scaling at ~ 0.7 $T_g$ (additional measurements were performed at $0.84T_g$ in the aligned sample).[12-14]

In the ZFCM protocol, samples are cooled from a temperature (140 K) well above the superspin glass transition temperature, to the measuring temperature, $T_m < T_g$, and held for an experimentally fixed amount of time $t_w$ (waiting time, or equivalently called, the system's 'age') in zero applied field. During $t_w$, superspins form correlated zones of various sizes. At $t_w$, an average sized correlated zone contains $N_s(t_w)$ correlated (super)spins, with a corresponding free energy barrier $B(N_s(t_w))$[3, 17]; i.e.,

$$t_w = \tau_o e^{\left(\frac{B(t_w)}{k_B T}\right)} \quad (1),$$

where $\tau_0$ is the microscopic flipping time of a (super)spin and depends on temperature as $\tau = \tau_o exp(E_a/k_B T)$. After a chosen waiting time $t_w$, a small magnetic field ($H$ = 0.15~8 Oe) is applied and the magnetization is recorded as a function of the probing time $t$, elapsed since the field change.

The values of $N_s$ are extracted from the ZFCM data following an empirical model developed for atomic spin glasses.[3] The magnetization relaxation toward the final value requires cooperative flipping of all (super)spins in a given correlated zone. Therefore, the magnetization relaxation rate spectrum, $S(log(t))$ (= $d(M/M_{FC})/d\log(t)$) is a qualitative representation of the size distribution of such zones.[17, 18] As average sized zones possess the relaxation time ~ $t_w$, $S$ becomes maximum near $t \sim t_w$ when $N_s(t_w)$ spins cooperatively flip, provided that the magnetic field is vanishingly small. When a larger magnetic field is applied, the Zeeman energy (magnetic field coupling to a group of correlated (super)spins) becomes non-negligible, and the barrier energy is reduced from $B(N_s(t_w))$ to $B(N_s(t_w)) - E_Z(H, N_s(t_w))$. Consequently, the relaxation rate reaches its maximum at $t = t_w^{eff} < t_w$. $t_w^{eff}$ is called the *effective age* of the system and is described as:

$$t_w^{eff} = \tau_o exp\{(B(t_w) - E_z(H, N_s(t_w))/k_B T\} \quad (2).$$

Combining equations 1 and 2, one can deduce the relation between the observed *effective age* ($t_w^{eff}$) and $E_z$ ($H$, $N_s$) of $N_s(t_w)$ correlated spins to:

$$ln(t_w^{eff}/t_w) = -(E_z(H, N_s(t_w))/k_B T \quad (3).$$

In atomic spin glasses, the form of $E_Z(H, N_s)$ was found to depend on the spin anisotropy nature. In one Ising-spin glass, $E_Z(H, N_s) \sim H$ was observed,[5] while in several Heisenberg spin glasses, $E_Z(H, N_s) \sim H^2$ was reported.[3-5] To account for these observations, the following empirical models were proposed:

$$E_Z(H) = \sqrt{N_s}\mu H; \text{ for Ising SG's, with relatively small } N_s \quad (4)$$

and

$$E_Z = N_s\chi_{FC}H^2 \text{ ; for Heisenberg SG's with macroscopically large } N_s \quad (5),$$

where $\mu$ is the magnetic moment and $\chi_{FC}$, the field cooled susceptibility per spin. While the extraction of $N_s$ from $E_z(H)$ is rather straightforward, the calculation of the correlation length, $\xi$, from $N_s$ is less palpable due to the fractal nature of the spatial correlations omnipresent in disordered systems such as spin glasses. To this end, Berthier *et al.* has determined numerically the fractal dimensionality of the "backbone spin structure", $d-\alpha$. In Ising and Heisenberg spin glasses these values correspond to ≈ $d-0.5$ and $d-1$ ($d = 3$ for 3D systems), respectively.[6,7] The simplest assumption is then to take $\xi/\xi_o = N_s^{d-\alpha}$ to deduce the correlation length from the experimentally determined values of $N_s$. $\xi_o$ is the average distance between two neighboring (super)spins.

## III. RESULTS AND DISCUSSIONS

As depicted in Figure 1, SPM magnetization of the aligned (textured) sample was about 3 times that of the random sample. While the glass transition temperature ($T_g$ ~ 69 K) was not affected by the anisotropy-axis alignment, there were appreciable changes in the SSG dynamics between the two systems. The notable differences between the two SSG's are as follows. a) The critical exponent values were found to increase slightly from $z\nu$ ≈ 7 in the random sample to ≈ 8.5 in the aligned sample (Figure 1, inset). b) Stronger cooling effect was observed in the aligned sample.[14] c) The field dependence of $t_w^{eff}$ (see Figure 2) was nearly quadratic in the random SSG, whereas it became linear in the aligned SSG. These contrasts closely mimic the reported differences between Heisenberg (weak cooling rate effect and $t_w^{eff} \sim H^2$) and Ising (strong cooling effect, larger $z\nu$, and $t_w^{eff} \sim H$) atomic spin glasses. With this analogy in mind, we have used the above mentioned empirical model for atomic Ising spin glasses (Eq. 4) to extract the number of correlated superspins in the aligned SSG sample.

The quadratic dependence of $E_z(H)$ in the random SSG, however, is only true at higher field values. At lower fields, the growth appears to be slower than $H^2$ (Fig. 2 bottom inset). We can interpret this slope change in the following manner. Randomly oriented or aligned, the superspins possess well defined anisotropy-axis. Therefore, for small values of $N_s$, the magnetization of randomly oriented superspins must follow $M(N_s) = \sqrt{(N_s/3)}\mu$, which contributes a linear term in $E_z(H)$, observable only at low fields. With increasing field strength and $N_s$, the magnetization will crossover to a macroscopic and extensive form, $N_s\chi_{FC}H$ and thus the quadratic term, $N_s\chi_{FC}H^2$, dominates the total Zeeman energy. The corresponding total $E_z$ in a random SSG is then expressed as,

$$E_z(H) = \sqrt{(N_s/3)}\mu H + N_s\chi_{FC}H^2 \qquad (6)$$

A quick verification reveals that for $N_s \sim 300$ (value previously reported in a maghemite frozen ferrofluid[12]) and the corresponding values of $\mu$ and $\chi_{FC}$ of the nanoparticles (same as those used in this study) the crossover from linear to quadratic dependence occurs at $H$ as low as a few Gausses.[19]

In Figure 3, the extracted values of $N_s$ of both aligned and random SSG's studied here along with the results from our previous report[11] are presented as a function of scaled time, $ln(t_w/\tau_o)T/T_g$, and compared to the experimental results in atomic spin glasses from various groups.[3-5] The scaling parameter is a direct consequence of the power law behavior of $\xi(t)$ as mentioned earlier. As can be seen from the figure, $N_s(t_w, T)$ data in two random SSG samples lie on the extension of data points collected from Heisenberg SG's. For the aligned SSG and the Ising spin glass data, a common growth curve maybe drawn (dotted line) to accommodate both data sets; however, the agreement is less evident than in the Heisenberg counterpart.

We now attempt to estimate the correlation length $\xi/\xi_o$ from $N_s$ in both aligned and random SSG's as well as in atomic spin glasses summarized in Figure 3. The fractal exponents used here are those introduced by Berthier and Young; namely $d-\alpha \approx 2.5$ (Ising SG and aligned SSG) and 2 (Heisenberg SG and random SSG).[6,7] One can see from Figure 4 that $\xi/\xi_o$ data of random SSG's position themselves nicely between the Heisenberg SG simulation curve and the experimental results within experimental error bars. It should be noted that in the simulation on Heisenberg SG's, a clear downward curvature was observed at low temperature and at large waiting times. Combined with the uncertainty associated with the fractal exponent values themselves, the qualitative agreement found between the experimentally extracted correlation lengths and the simulation results must be regarded with precaution. However, it should not be an overstatement to say that the Heisenberg simulation

47curves and the experimental results are in quantitative agreement. Further, the correlation length growths in randomly oriented SSG's appear to follow the same physical law as that of atomic spin glasses. The correlation lengths in the Ising spin glass sample fall upon the extension of the Ising simulation curves, which are found to be nearly temperature independent and follow $\xi \sim (t_w)^{zT/T_g}$. In the aligned SSG, the correlation length appears to fall somewhere between the Heisenberg and Ising type dynamics. This may not be too surprising considering that magnetic nanoparticles do not possess infinite anisotropy energy, and therefore, our aligned sample may still be far from a true Ising superspin glass system.

## IV. CONCLUSIONS

We have extracted the growing number of dynamically correlated superspins in the SSG state of textured (aligned) and non-textured (randomly oriented) frozen ferrofluids via ZFCM relaxation measurements. The number of correlated spins, $N_s$, in randomly oriented superspin glasses lies on the extension of the general curve found in Heisenberg spin glasses[3-5]. The corresponding correlation lengths of random SSG and Heisenberg-like SG's estimated using the fractal exponent suggested by Berthier and Young are in a quantitative agreement with the numerical simulations on Heisenberg SG's by the same authors.[6,7]

This work demonstrates the usefulness of interacting magnetic nanoparticle systems to revisit the physics of spin glass by virtue of their tunable physical parameters. With a right combination of particle size (tunes $\tau_o(T)$ and $T_B$) and concentration (tunes $T_g$) one can hope to fully bridge the gap between the experiments and the numerical simulations left behind by decades of research in atomic spin glasses.

**ACKNOWLEDGEMENTS**

S.N. and E.V. thank Prof. S. Miyashita and Prof. H. Yoshino for insightful discussions and suggestions.

Figure captions:

Figure 1: ZFC/FC of aligned (textured) and random (non-textured) ferrofluids measured at H = 1 Oe. A sudden drop in the magnetization of aligned sample near 200 K indicates the onset of the melting of glycerol. Inset: determination of the critical exponents on both ferrofluids obtained from ac susceptibility measurements.

Figure 2: The effective waiting times in ZFCM in aligned (top) and random (bottom) SSG samples at $T = 0.7\ T_g$ as a function of $H$ and $H^2$, respectively. The insets show the log-log representation of Zeeman energy as a function of $H/H_o$ with $H_o = 1$ Oe in the aligned SSG and = 1.4 Oe in the random SSG.

Figure 3: (color online) $N_s$ ($t_w$,$T$) extracted using Equations 4 and 6, plotted against $T/T_g ln(t_w/\tau_o^*)$ compared to the experimental results reported in atomic spin glasses.[3-5] $\tau_o^*$ was calculated according to $\tau_o^* = \tau_o\ exp(E_a/k_B T)$ with $\tau_o = 10^{-9}$ sec and $E_a = 640$ K. $N_s(t, T)$ in random SSG's (this and previous work[12]) coincide with the scaling curve found among Heisenberg atomic spin glasses within experimental uncertainties.

Figure 4: (color online) Correlation lengths ($\xi/\xi_o$) calculated from $N_s$ in aligned and random SSG's as well as in atomic SG's[3-5] (experimental) using $\alpha = 0.5$ (Ising SG and aligned SSG) and 1.0 (Heisenberg SG and random SSG). The values are plotted as a function of scaled time (same symbols as in Fig. 3) and compared to the numerical simulations.[6,7] Symbols: Stars - Ising SG (numerical) at $T = T_g$ and $0.5T_g$, solid triangles -from right to left, Heisenberg SG (numerical) at $T = T_g$, $0.875T_g$, $0.75T_g$, $0.5T_g$ and $0.25T_g$, solid circles - Atomic SG's (experimental), and octagons – SSG's (this work).

Figure 1

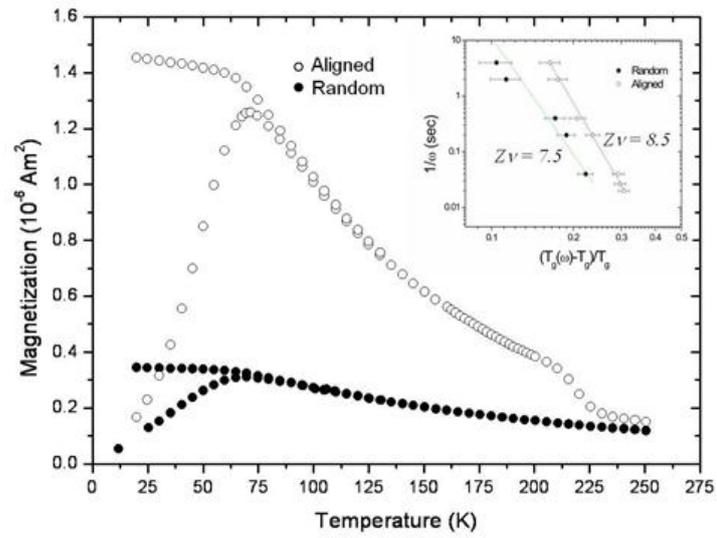

Figure 2:

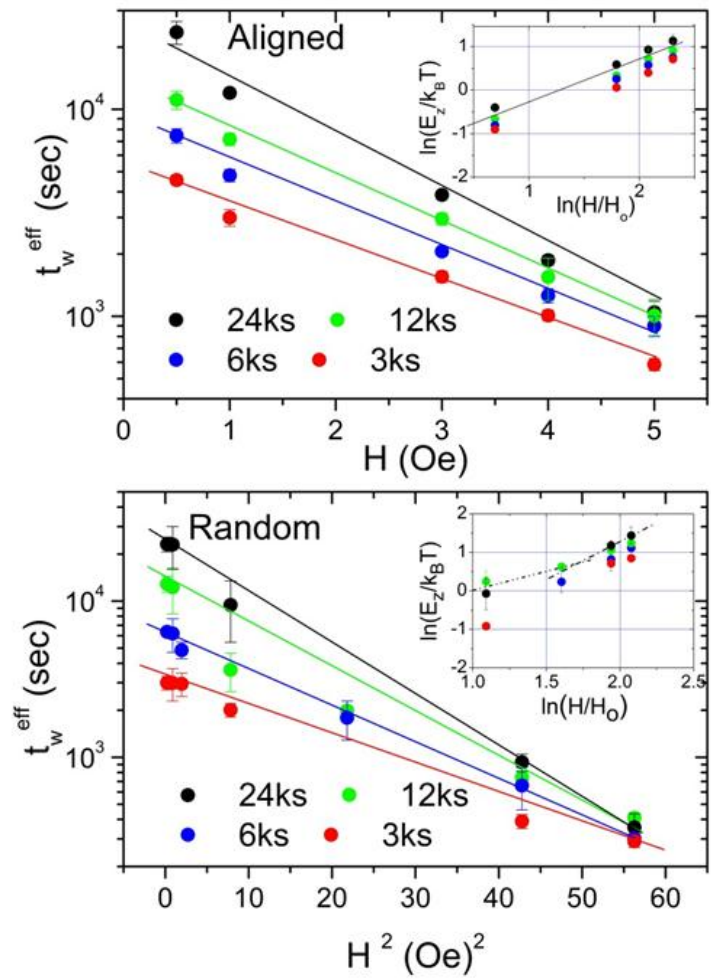

12Figure 3

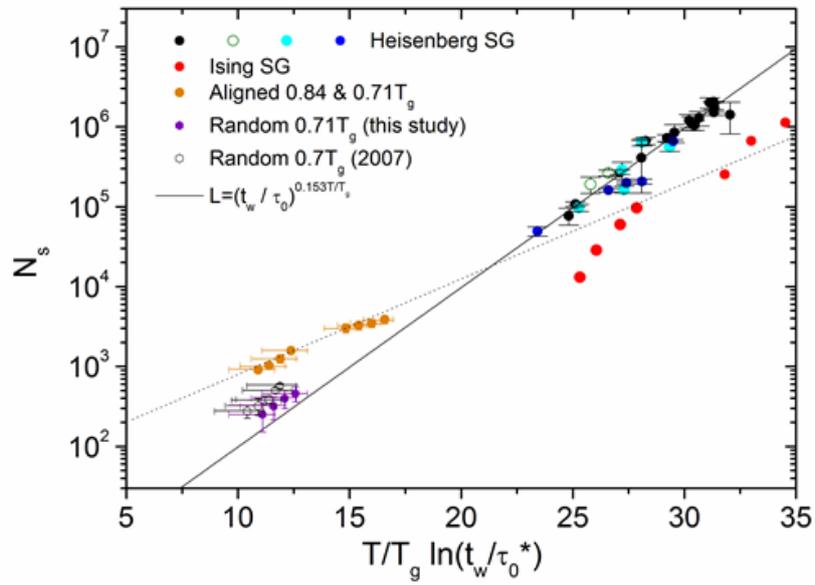

Figure 4

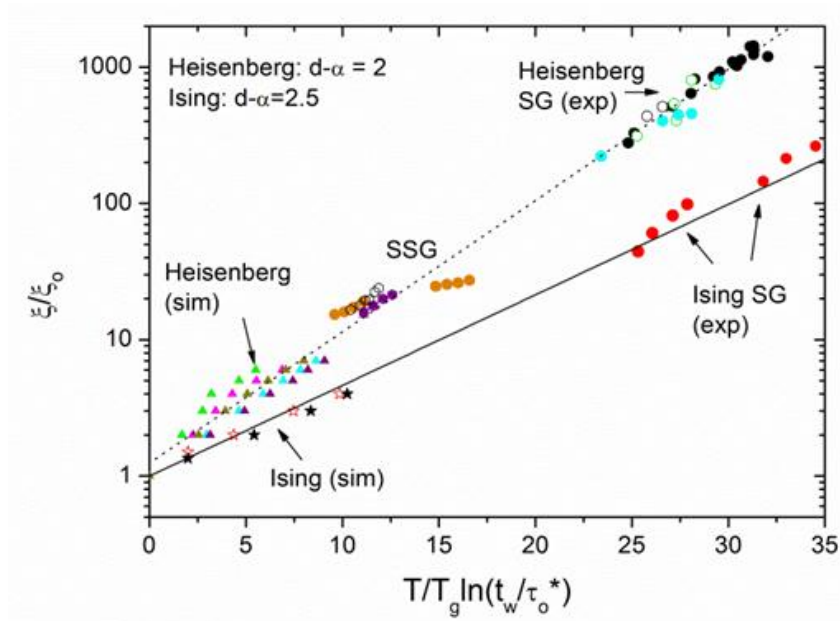

12